\documentclass[AMA,STIX1COL]{WileyNJD-v2}

\articletype{Research Article}

\received{}
\revised{}
\accepted{}

\newcommand\independent{\protect\mathpalette{\protect\independenT}{\perp}}
\def\independenT#1#2{\mathrel{\rlap{$#1#2$}\mkern2mu{#1#2}}}

\DeclareMathOperator*{\argmax}{arg\,max}

\renewcommand{\thesection}{\arabic{section}}%

\begin{document}

\title{Transporting Experimental Results with Entropy Balancing}

\author[1]{Kevin P. Josey}
\author[2]{Seth A. Berkowitz}
\author[1]{Debashis Ghosh}
\author[3,4]{Sridharan Raghavan*}

\authormark{Josey \textsc{et al}.}

\address[1]{\orgdiv{Department of Biostatistics and Informatics}, \orgname{Colorado School of Public Health, University of Colorado Anschutz Medical Campus}, \orgaddress{\state{Colorado}, \country{USA}}}
\address[2]{\orgdiv{Division of General Medicine and Clinical Epidemiology}, \orgname{University of North Carolina at Chapel Hill School of Medicine}, \orgaddress{\state{North Carolina}, \country{USA}}}
\address[3]{\orgname{Rocky Mountain Regional VA Medical Center}, \orgaddress{\state{Colorado}, \country{USA}}}
\address[4]{\orgdiv{Division of Hospital Medicine}, \orgname{University of Colorado School of Medicine}, \orgaddress{\state{Colorado}, \country{USA}}}

\corres{*Sridharan Raghavan, Rocky Mountain Regional VA Medical Center, \email{sridharan.raghavan@va.gov}}

\presentaddress{1700 N Wheeling St, Aurora, CO 80045}

\abstract[Summary]{We show how entropy balancing can be used for transporting experimental treatment effects from a trial population onto a target population. This method is doubly-robust in the sense that if either the outcome model or the probability of trial participation is correctly specified, then the estimate of the target population average treatment effect is consistent. Furthermore, we only require the sample moments of the effect modifiers drawn from the target population to consistently estimate the target population average treatment effect. We compared the finite-sample performance of entropy balancing with several alternative methods for transporting treatment effects between populations. Entropy balancing techniques are efficient and robust to violations of model misspecification. We also examine the results of our proposed method in an applied analysis of the Action to Control Cardiovascular Risk in Diabetes Blood Pressure (ACCORD-BP) trial transported to a sample of US adults with diabetes taken from the National Health and Nutrition Examination Survey (NHANES) cohort.}


\jnlcitation{}

\maketitle

\footnotetext{}

\section{Introduction}

In a randomized controlled trial (RCT), the population from which the sample is collected, the \textit{trial population}, often differs from the population of interest, the \textit{target population}. This scenario becomes problematic when the true causal effect is heterogeneous, implying the existence of effect modifying covariates - \textit{effect modifiers} - which alter the average treatment effect. If the distribution of the effect modifiers is different in the trial and target populations, the average treatment effect observed in the trial will likely differ from what would be observed within the target population, limiting the conclusions that can be drawn from an otherwise well-designed study. It is worth noting that effect modifiers are specific to the scale of the target estimand. Throughout, we will refer to effect modifiers as variables that modify the average treatment effect which we will define later on.

\begin{figure}
	\centering
	\includegraphics[scale = 1]{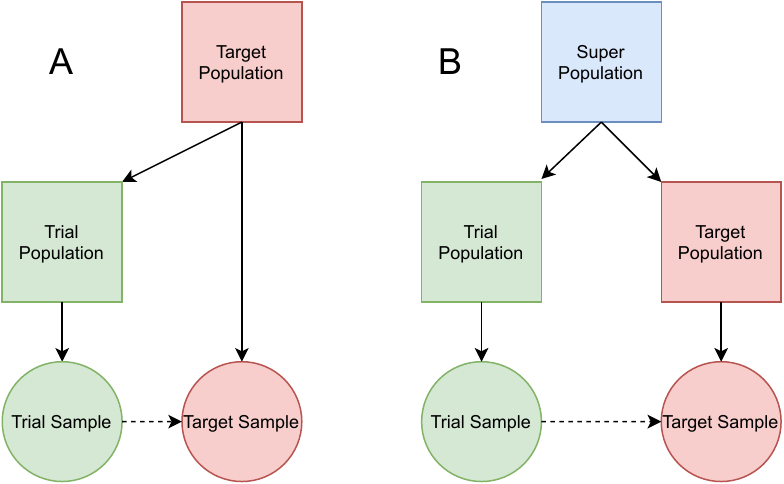}
	\caption{Square nodes represent populations whereas circular nodes represent samples. The solid arrow represents a subsetting of the origin node. The dashed line represents the process of generalizability (A) and transportability (B).}\label{generalize}
\end{figure}

The recent literature on the subject of transportability is divided into two scenarios determined by the nature of the trial and target populations, and the desired causal estimand. If the trial population is nested within the target population, we can extend the results of an RCT using a sample from the target population in a process called \textit{generalizability}. If the target and trial populations are subpopulations drawn from some super population, then the problem is one of \textit{transportability}. Figure \ref{generalize} provides a diagram relating the data to the corresponding populations in both the generalizeability and transportability problems. Note that for generalizability, the trial population is a subpopulation of the target population, while in transportability the target and trial populations are not nested. We will discuss the difference between these two scenarios in more detail in Section \ref{assumptions}. The work herein, however, will focus primarily on the issue of transportability.

Some articles have approached the problem of transportability from the setting in which the investigator is provided the individual-level data from the trial population along with individual-level covariate data from the target population.\cite{rudolph_robust_2017} Another setting provides the individual-level data from the trial population, but only the covariate sample moments (e.g., the mean and standard deviation) from the target population, which can often be found in a so-called Table 1 throughout the medical literature.\cite{signorovitch_comparative_2010} One property that is often sought while developing estimators for causal inference is called double-robustness.\cite{bang_doubly_2005} In the context of transporting experimental results, this means that if either the probability of trial participation or the outcome model are correctly specified, then the resulting average treatment effect estimator is consistent.

We propose using entropy balancing to solve transportability problems. The procedure we propose builds upon several other causal effect estimators which employ convex optimization techniques to estimate a vector of \textit{sampling weights}.\cite{signorovitch_comparative_2010, hartman_sample_2015, zhang_new_2016, phillippo_methods_2018} These sampling weights would otherwise be uniform if the RCT data were randomly sampled from the target population. The literature on convex optimization in the context of causal inference has abounded in recent years.\cite{hainmueller_entropy_2012, imai_covariate_2014, wang_minimal_2020} Rather than using these methods to exactly balance the covariate distribution between the treated and control units within an observational study, convex optimization techniques applied to transportability can be used to estimate weights which balance the covariate distribution of the trial participants and non-participants. Entropy balancing is flexible in that it can be applied both when the complete individual-level covariate data are provided as well as when only the covariate sample moments of the target population are provided, such as what might appear in the Table 1 of a clinical paper. Furthermore, the specific entropy balancing procedure we develop can be shown to be doubly-robust for estimating the target population average treatment effect given the complete individual-level covariate data in the context of transportability.

The entropy balancing solution we propose is also considered as a solution for indirectly comparing experimental results from two separate randomized trials with an anchored treatment arm, a problem not too dissimilar from that of transportability. However, transportability as we describe it does not require a second randomized trial.  The sample drawn from the target population, which is subsequently used in our transportability formulations, does not require any information about the outcome or treatment assignment. Moreover, based on new results identified in the indirect comparison literature, we can relax a rather strong assumption about the nature of the potential outcomes typically made in the transportability literature.\cite{phillippo_methods_2018} In addition, we take a more comprehensive view of entropy balancing through the lens of causal inference, motivating this work through the potential outcomes framework and describe several properties about entropy balancing for transportability that are sought for estimators in the causal inference literature. This includes the property of double-robustness, semiparametric efficiency, and considerations between finite-sample and superpopulation settings. We also compare the entropy balancing approach with several other methods in an effort to showcase the strengths of doubly-robust estimators more generally in transportability problems.

The contents of the article are as follows. In Section \ref{setting} we define the notation, setting, and assumptions necessary for transporting experimental results between populations and describe several existing methods for transportation, including two methods that can be applied in the setting where we are given only the sample covariate moments of the target population and two methods that require individual-level covariate data from the target population, one of which is doubly-robust. In Section \ref{entropy}, we introduce entropy balancing and describe the difference between conducting inference upon the target population average treatment effect versus the target sample average treatment effect. In Section \ref{simulation} we compare the five methods considered in Sections \ref{setting} and \ref{entropy} using numerical studies. We also illustrate through a secondary simulation how entropy balancing and other methods that do not require individual-level data from the target population only allow for inference upon the target sample average treatment effect and not the target population average treatment effect. In Section \ref{illustrative} we compare entropy balancing and inverse odds of sampling weights in a real-data example: transporting results from a clinical trial of blood pressure treatment intensity in diabetes patients to a representative sample of the US population. Section \ref{discussion} concludes with a discussion.

\section{Setting and Preliminaries}\label{setting}

\subsection{Notation and Potential Outcomes}\label{notation}

Suppose we have two random samples from different populations. For independent sampling units $i = 1,2,\ldots,n$, let $S_i \in \{0,1\}$ denote a random sampling indicator. Indexed by $\{i : S_i = 1\}$, the \textit{trial sample} evaluates the efficacy of some treatment on the trial population. The second sample is randomly selected from the target population and indexed by $\{i : S_i = 0\}$. We refer to this sample as the \textit{target sample}. We denote $n_1 = \sum_{i = 1}^n S_i$, $n_0 = \sum_{i = 1}^n (1 - S_i)$, and $n = n_1 + n_0$. Both $\mathbb{E}(\cdot)$ and $\Pr\{\cdot\}$ will be evaluated over the \textit{superpopulation} which is the combined trial and target population.

For $i = 1,2\ldots, n$, let $\mathbf{X}_{i} \in \mathcal{X}$ denote a vector of measured baseline (i.e. pretreatment) covariates. For $i\in\{i:S_i = 1\}$, let $Y_{i} \in \Re$ denote the real valued outcome, and $Z_i \in \{0,1\}$ denote the random treatment assignment. We assume throughout that $\mathbf{X}_i$ contains an intercept term. The probability density function for $\mathbf{X}_i$ is denoted $f(\mathbf{x}_i)$ for $\mathbf{x}_i \in \mathcal{X}$. Indexed by $j = 1,2,\ldots,m$, we denote the vector-valued balance function $\mathbf{c}(\mathbf{X}_i) \equiv [c_1(\mathbf{X}_i), c_2(\mathbf{X}_i), \ldots, c_j(\mathbf{X}_i), \ldots, c_m(\mathbf{X}_i)]^T$, which are the effect modifiers included into the models of $S_i$ and $Y_i$ along with $Z_i$. We suppose $c_1(\mathbf{X}_i) = 1$ for all $i = 1,2,\ldots,n$.  Some examples for $c_j(\mathbf{X}_i)$, $j = 2,3,\ldots,m$ include polynomial transformations of the covariates and interaction terms - anything that might modify the effect of the treatment on the outcome. It is common practice to balance the covariates as well as the variance (i.e. second moments) of the covariates when more intimate knowledge about the effect modifying process is unknown.\cite{signorovitch_comparative_2010}

We employ the potential outcomes framework for a binary treatment.\cite{rubin_estimating_1974}  This framework allows us to construct the observed outcome in terms of the factual and counterfactual variables $Y_{i}(0)$ and $Y_{i}(1)$, $i = 1,2,\ldots,n$. $Y_{i}(0)$ and $Y_{i}(1)$ correspond to each unit's outcomes when $Z_{i} = 1$ and $Z_{i} = 0$, respectively. The observed responses are then defined as $Y_{i} \equiv Z_{i} Y_{i}(1) + (1 - Z_{i})Y_{i}(0)$. The potential outcomes framework also allows us to define the target population average treatment effect, $\tau_{\text{TATE}} \equiv \mathbb{E}[Y_i(1) - Y_i(0)|S_i = 0]$ and the target sample average treatment effect \[ \tau_{\text{TSATE}} \equiv \frac{1}{n_0}\sum_{\{i : S_i = 0 \}} \left[Y_i(1) - Y_i(0)\right]. \] The target sample average treatment effect only concerns the effects among units within the target sample whereas the target population average treatment effect concerns the average effect for all units that make up the target population.\cite{imai_misunderstandings_2008} We also define $\rho(\mathbf{X}_i) \equiv \Pr\{S_i = 1|\mathbf{X}_i\}$ and $\pi \equiv \Pr\{Z_i = 1|\mathbf{X}_i\}$. Recall that in an RCT, $\pi \in (0,1)$ should be independent of $\mathbf{X}_i$. Another way to write $\tau_{\text{TATE}}$ is \[ \tau_{\text{TATE}} \equiv \frac{ \mathbb{E}\left\{[1 - \rho(\mathbf{X}_i)][Y_i(1) - Y_i(0)]\right\}}{\mathbb{E}[1 - \rho(\mathbf{X}_i)]}. \] This alternative definition identifies the target population average treatment effect as a type of weighted average treatment effect. A corollary to Theorem 4 of Hirano \textit{et al}. can be used to derive the semiparametric efficiency bound for any estimator of $\tau_{\text{TATE}}$ as 
\begin{equation}\label{semi}
\Sigma_{\text{semi}} \equiv \frac{\mathbb{E}\left([1 - \rho(\mathbf{X}_i)]^2 \left\{\frac{\mathbb{V}[Y_i(1)|\mathbf{X}_i]}{\pi} + \frac{\mathbb{V}[Y_i(0)|\mathbf{X}_i]}{1 - \pi} + \left[\tau_{\text{TATE}}(\mathbf{X}_i) - \tau_{\text{TATE}} \right]^2 \right\} \right)}{\mathbb{E}[1 - \rho(\mathbf{X}_i)]^2}
\end{equation} where $\tau_{\text{TATE}}(\mathbf{X}_i) \equiv \mathbb{E}[Y_i(1) - Y_i(0)|\mathbf{X}_i, S_i = 0]$.\cite{hirano_efficient_2003} This setup allows us to utilize the asymptotic results about weighted average treatment effects derived previously using convex optimization techniques such as those employed by entropy balancing.\cite{chan_globally_2016}

We denote the population moments of the target covariate distribution as $\mathbb{E}[\mathbf{c}(\mathbf{X}_i)|S_i = 0] = \boldsymbol{\theta}_0$. For much of this paper, we will describe methods for transporting experimental results which weight the responses $\mathbf{Y}_i$ for $i \in \{i:S_i = 1\}$ so that the weighted trial sample moments are the same as the population moments of the target population.\cite{deville_calibration_1992} We will denote the sample weights as $\boldsymbol{\gamma} \equiv (\gamma_1, \gamma_2, \ldots, \gamma_{n_1})$. Since $\boldsymbol{\theta}_0$ is usually unknown, we will need to make use of the estimator $\hat{\boldsymbol{\theta}}_0 \equiv n_0^{-1}\sum_{\{i:S_i = 0\}} \mathbf{c}(\mathbf{X}_i)$. In practice, we usually set $\mathbf{c}(\mathbf{X}_i) = \mathbf{X}_i$ unless more is known about the data generating mechanisms. In those cases, $\hat{\boldsymbol{\theta}}_0$ typically appears in the so-called Table 1 of many publications.

\subsection{Assumptions for Transportability}\label{assumptions}

The following assumptions facilitate our ability to transport experimental results onto a target population. These assumptions represent the necessary conditions required for the transporting experimental results. They are also adapted from similar articles on the subject. \cite{pearl_external_2014, rudolph_robust_2017, dahabreh_extending_2020} Furthermore, we invoke the stable unit treatment value, which comprises the no interference and consistency assumptions.

\begin{assumption}[Mean Difference Exchangeability]\label{exchange}
The target population average treatment effect conditional on the baseline covariates is exchangeable between samples: \[ \mathbb{E}[Y_{i}(1) - Y_{i}(0)|\mathbf{X}_i, S_i = 1] = \mathbb{E}[Y_{i}(1) - Y_{i}(0)|\mathbf{X}_i, S_i = 0] = \mathbb{E}[Y_{i}(1) - Y_{i}(0)|\mathbf{X}_i]. \]
\end{assumption}

\begin{assumption}[Sampling Positivity]\label{positivity}
The probability of trial participation, conditioned on the baseline covariates necessary to ensure Assumption \ref{exchange}, is bounded away from zero and one: \[ 0 < \Pr\{S_i = 1|\mathbf{X}_i = \mathbf{x}_i \} < 1 \ \text{for all} \ \mathbf{x}_i \in \mathcal{X} \ \text{where} \  f(\mathbf{x}_i|S_i = 0) > 0. \] 
\end{assumption}

\begin{assumption}[Strongly Ignorable Treatment Assignment]\label{sita}
The potential outcomes among the trial participants are independent of the treatment assignment given $\mathbf{X}_i$: \[ [Y_{i}(0), Y_{i}(1)]^T \independent Z_{i} | \mathbf{X}_{i} \ \text{for all} \ i \in \{i:S_i = 1\}. \]
\end{assumption}

Assumption \ref{sita} is a standard assumption in the potential outcomes literature.\cite{rosenbaum_central_1983} This assumption can be further simplified in an RCT setting to assume \[ [Y_{i}(0), Y_{i}(1)]^T \independent Z_{i} \ \text{for all} \ i \in \{i:S_i = 1\} \] since there should be no association between the treatment assignment and the covariates. The covariate imbalance that requires amelioration in transportability instead appears between $\mathbf{X}_i$ and $S_i$. 

As noted previously in the Introduction, there are subtle distinctions between generalizability and transportability. The main difference occurs with the causal estimand of interest. In transportability, the target estimand is $\tau_{\text{TATE}}$. For generalizability, the causal estimand of interest is $\tau_{\text{ATE}} \equiv \mathbb{E}\left[Y_i(1) - Y_i(0) \right]$. This is on account of the trial population being nested within the target population, so the superpopulation and the target population are identical. Under our notation, generalizability further assumes that the units $\{i:S_i = 0\}$ are sampled from the target population and the complement of the trial population. As a result, we would need to rewrite Assumption \ref{positivity} for generalizability to state \[ 0 < \Pr\{S_i = 1|\mathbf{X}_i = \mathbf{x}_i \} < 1 \ \text{for all} \ \mathbf{x}_i \in \mathcal{X} \ \text{where} \  f(\mathbf{x}_i) > 0. \] We avoid this setup to the problem and instead focus on methods developed for transportability and inference on $\tau_{\text{TATE}}$.

In addition to Assumptions \ref{exchange}-\ref{sita}, we require the following assumptions to establish the double-robustness property of entropy balancing. We will show that if either assumption is met, then the entropy balancing methods introduced in Section \ref{entropy} will be consistent for the target population average treatment effect. We also use these assumptions to establish consistency for some of the other methods we describe in Section \ref{previous} when standard regression methods are employed. Note that an underlying requirement implied by these two assumptions is that there is no unnmeasured effect modification present given the known balance functions. If an effect modifier is missing, then any estimator we present will likely produce biased estimates of the target population average treatment effect.

\begin{assumption}[Conditional Linearity]\label{linear}
The expected value of the potential outcomes, conditioned on $\mathbf{\mathbf{X}_i}$, is linear across the span of the covariates. That is $\mathbb{E}[Y_{i}(1) - Y_{i}(0)|\mathbf{X}_i] = \mathbf{c}(\mathbf{X}_i)^{T}\boldsymbol{\alpha}$ and $\mathbb{E}[Y_{i}(0)|\mathbf{X}_i] = \mathbf{c}(\mathbf{X}_i)^{T}\boldsymbol{\beta}$ for all $i = 1,2,\ldots,n$ and $\boldsymbol{\alpha},\boldsymbol{\beta} \in \Re^m$.
\end{assumption}

\begin{assumption}[Linear Conditional Log-Odds]\label{odds}
The log-odds of trial participation are linear across the span of the covariates. That is $\text{logit}[\rho(\mathbf{X}_i)] = \mathbf{c}(\mathbf{X}_i)^{T} \boldsymbol{\lambda}$ for all $i = 1,2,\ldots,n$ and some $\boldsymbol{\lambda} \in \Re^m$.
\end{assumption}

Assumption \ref{exchange} is substantially relaxed from what appears in more recent literature.\cite{rudolph_robust_2017, dahabreh_extending_2020} These other articles require the expected value of the potential outcomes to be exchangeable between populations: 
\begin{equation}\label{absolute}
    \mathbb{E}[Y_{i}(1)|\mathbf{X}_i, S_i = 1] = \mathbb{E}[Y_{i}(1)|\mathbf{X}_i, S_i = 0] \quad \text{and} \quad \mathbb{E}[Y_{i}(0)|\mathbf{X}_i, S_i = 1] = \mathbb{E}[Y_{i}(0)|\mathbf{X}_i, S_i = 0].
\end{equation}  The indirect comparison literature refers to this assumption as the conditional constancy of \textit{absolute} effects whereas Assumption \ref{exchange} is commonly referred to as the conditional constancy of \textit{relative} effects. Whereas the conditional constancy of absolute effects requires adjustment for all prognostic and effect modifying covariates, the conditional constancy of relative effects only requires adjustments for the effect modifiers. Note that the much stronger conditional constancy of absolute effects assumption is enforced in  Assumption \ref{linear}. However, the less stringent Assumption \ref{exchange}, in addition to Assumptions \ref{positivity} and \ref{sita}, is all that is required to obtain consistent estimates when Assumption \ref{odds} holds. This assumption relaxation result can be made using arguments of anchored indirect comparisons.\cite{phillippo_methods_2018} Suppose the target sample is a randomized control trial comparing $Z = 1$ with $Z = 0$, similar to the data contained within the trial sample, only we do not observe either the outcome or the treatment assignment. Then this target ``trial" is figuratively anchored by both treatment groups. If the target ``trial" were to be conducted, and the outcome and treatment data were collected, then a resulting indirect comparison estimator should yield estimates of zero as both ``trials" examine the same estimand.\cite{phillippo_equivalence_2020} This is precisely what transportability methods are targeting – what the causal effect would be if the trial were conducted on a different population. Only requiring the conditional constancy of relative effects assumption versus the conditional constancy of absolute effects assumption adds incentive to focus on correctly specifying the sampling model over the outcome model through the entropy balancing techniques that we will describe in Section \ref{entropy}. 

\subsection{Alternative Methods for Transportability}\label{previous}

In this section we present four different methods for transporting experimental results to estimate $\tau_{\text{TATE}}$. For each method, we assume we are given Assumptions \ref{exchange}-\ref{sita}. The first method weights responses of the trial sample with the inverse odds of sampling.\cite{westreich_transportability_2017} Define the inverse odds of sampling weights as \[ \hat{\gamma}^{\text{PS}}_i = \begin{cases} \frac{1 - \hat{\rho}(\mathbf{X}_i)}{\hat{\rho}(\mathbf{X}_i)\hat{\pi}}, & \text{when} \ S_i = 1, Z_i = 1 \\ \frac{1 - \hat{\rho}(\mathbf{X}_i)}{\hat{\rho}(\mathbf{X}_i)(1 - \hat{\pi})}, & \text{when} \ S_i = 1, Z_i = 0 \\ 0, & \text{when} \ S_i = 0 \end{cases} \] where $\hat{\pi}$ is a consistent estimator of the probability of treatment and $\hat{\rho}(\mathbf{X}_i)$ is a consistent estimator of the probability of trial participation. The target population average treatment effect is then estimated by computing \[ \hat{\tau}_{\text{IOSW}} = \sum_{\{i:S_i = 1\}} \frac{\hat{\gamma}^{\text{PS}}_iZ_iY_i}{\sum_{\{i:S_i = 1\}} \hat{\gamma}^{\text{PS}}_iZ_i} - \sum_{\{i:S_i = 1\}} \frac{\hat{\gamma}^{\text{PS}}_i(1 - Z_i)Y_i}{\sum_{\{i:S_i = 1\}} \hat{\gamma}^{\text{PS}}_i(1 - Z_i)}. \] If Assumption \ref{odds} is given, we may use logistic regression to consistently estimate $\hat{\rho}(\mathbf{X}_i)$. A consistent estimator for $\rho(\mathbf{X}_i)$ by extension renders $\hat{\tau}_{\text{IOSW}}$ consistent for $\tau_{\text{TATE}}$.

Given the extended conditional constancy of absolute effects assumption in (\ref{absolute}), another proposed solution is to find a consistent estimator of the conditional means for the potential outcomes with the sample data; $\mu_1(\mathbf{X}_i) \equiv \mathbb{E}[Y_i(1)|\mathbf{X}_i, S_i = 1]$ and $\mu_0(\mathbf{X}_i) \equiv \mathbb{E}[Y_i(0)|\mathbf{X}_i, S_i = 1]$. We will refer to this method as the outcome modeling (OM) approach. The consistent estimators are denoted as $\hat{\mu}_1(\mathbf{X}_i)$ and $\hat{\mu}_0(\mathbf{X}_i)$, respectively. Under Assumption \ref{exchange}, $\tau_{\text{TATE}}$ can be estimated by solving for \[ \hat{\tau}_{\text{OM}} = \frac{1}{n_0} \sum_{\{i: S_i = 0\}} \left[\hat{\mu}_1(\mathbf{X}_i) - \hat{\mu}_0(\mathbf{X}_i)\right]. \] In the causal inference literature, this method follows the framework for computing causal effects known as g-computation.\cite{robins_new_1986} If Assumption \ref{linear} is given, we can estimate $\tau_{\text{TATE}}$ with the OM approach if we are only given $\hat{\boldsymbol{\theta}}_0$ instead of $\mathbf{X}_i$ for all $i \in \{i:S_i = 0\}$. To do so, we would regress $Y_i$ onto $\mathbf{c}(\mathbf{X}_i)$ for all $\{i:S_i = 1, Z_i = 1\}$ and $\{i:S_i = 1, Z_i = 0\}$ to get $\hat{\boldsymbol{\alpha}}$ and $\hat{\boldsymbol{\beta}}$, respectively. We then compute \[ \hat{\tau}_{\text{OM}} = \hat{\boldsymbol{\theta}}^T_0\left(\hat{\boldsymbol{\alpha}} - \hat{\boldsymbol{\beta}}\right) \] where $\hat{\boldsymbol{\alpha}}$ and $\hat{\boldsymbol{\beta}}$ can be fit with ordinary least squares.

The OM approach and the inverse odds of sampling weights may be combined into a so called doubly-robust estimator. A doubly robust (DR) estimator combines estimators of the model components, in this case the model for $[Y_i(1), Y_i(0)]$ and $S_i$, as to be consistent when at most one model is misspecified. The conventional doubly-robust estimator for a binary treatment was first proposed as a semiparametric technique for missing-data problems.\cite{robins_estimation_1994} There have been extensive modifications to this conventional doubly-robust estimator, including alterations for transporting experimental results of a binary treatment.\cite{zhang_new_2016} Using the same notation outlined for the outcome model approach and the inverse odds of sampling weights, the target population average treatment effect can be estimated by solving for
\begin{equation}\label{dr}
\hat{\tau}_{\text{DR}} = \sum_{\{i:S_i = 1\}} \frac{\hat{\gamma}^{\text{PS}}_iZ_i[Y_i - \hat{\mu}_1(\mathbf{X}_i)]}{\sum_{\{i:S_i = 1\}} \hat{\gamma}^{\text{PS}}_iZ_i} - \sum_{\{i:S_i = 1\}} \frac{\hat{\gamma}^{\text{PS}}_i(1 - Z_i)[Y_i - \hat{\mu}_0(\mathbf{X}_i)]}{\sum_{\{i:S_i = 1\}} \hat{\gamma}^{\text{PS}}_i(1 - Z_i)} + \frac{1}{n_0} \sum_{\{i: S_i = 0\}} \left[\hat{\mu}_1(\mathbf{X}_i) - \hat{\mu}_0(\mathbf{X}_i)\right].
\end{equation}
It is easy to see that if $\hat{\mu}_0(\mathbf{X}_i)$ and $\hat{\mu}_1(\mathbf{X}_i)$ are consistent for $\mu_0(\mathbf{X}_i)$ and $\mu_1(\mathbf{X}_i)$, respectively, then the last summation in (\ref{dr}) is consistent for the target sample average treatment effect (and therefore also for the target population average treatment effect) while the first two summations converge in probability to zero. Similarly, if $\hat{\gamma}^{\text{PS}}$ is consistent for $\rho(\mathbf{X}_i)[1 - \rho(\mathbf{X}_i)]^{-1}$, then the first two summations of (\ref{dr}) will consistently estimate the negative bias induced by the last summation.\cite{dahabreh_extending_2020}

Another doubly-robust estimator closely related to the augmented estimator in (\ref{dr}) uses targeted maximum likelihood estimation (TMLE).\cite{laan_targeted_2006} TMLE is a popular choice among causal practitioners due to its flexibility for estimating a variety of different estimands, including the target population average treatment effect.\cite{rudolph_robust_2017} For transportability, the targeted maximum likelihood estimator is constructed as follows. First, the initial estimates of $\hat{\mu}_1(\mathbf{X}_i)$ and $\hat{\mu}_0(\mathbf{X}_i)$ are fit using the trial sample data. We then update the predictions of the potential outcomes on the trial sample with 
\begin{equation}\label{offset} 
\begin{split}
\tilde{\mu}_0(\mathbf{X}_i) &= \hat{\mu}_0(\mathbf{X}_i) + \hat{\epsilon}_0 (1 - Z_i) \hat{\gamma}^{\text{PS}}_i \enskip \\
\tilde{\mu}_1(\mathbf{X}_i) &= \hat{\mu}_1(\mathbf{X}_i) + \hat{\epsilon}_1 Z_i \hat{\gamma}^{\text{PS}}_i.
\end{split}
\end{equation} 
The estimates of $\epsilon_0$ and $\epsilon_1$ are obtained by regressing the outcome $Y_i$ onto the clever covariates - $Z_i \hat{\gamma}^{\text{PS}}_i$ and $(1 - Z_i)\hat{\gamma}^{\text{PS}}_i$ - with $\hat{\mu}_0(\mathbf{X}_i)$ and $\hat{\mu}_0(\mathbf{X}_i)$ serving as offsets for all $i \in \{i:S_i = 1\}$. The estimator of $\tau_{\text{TATE}}$ under the TMLE framework solves for 
\begin{equation}\label{tmle}
\hat{\tau}_{\text{TMLE}} = \frac{1}{n_0} \sum_{\{i: S_i = 0\}} \left[\tilde{\mu}_1(\mathbf{X}_i) - \tilde{\mu}_0(\mathbf{X}_i) \right]
\end{equation} 
in a similar manner to the OM approach. Equation (\ref{tmle}) is doubly-robust for estimating $\tau_{\text{TATE}}$ in the sense that if either the sampling model or the potential outcomes models are consistent, then $\hat{\tau}_{\text{TMLE}}$ is also consistent.\cite{rudolph_robust_2017} TMLE also requires individual-level covariate data to estimate some of the components in (\ref{offset}) and (\ref{tmle}).

For the DR and TMLE methods, it is unclear to us whether the more relaxed Assumption \ref{exchange} remains applicable in cases where the sampling model is correctly specified. Note that both these methods are heavily geared toward the outcome regression model being correctly specified. To avoid any distraction from this potential discrepancy, we ensure the conditional constancy of absolute effects assumption is satisfied in the simulation study found in Section \ref{sim_1}. Furthermore, we will compare the entropy balancing methods described in the next section with IOSW alone in the data analysis found in Section \ref{illustrative} as the conditional constancy of absolute effects assumption cannot be guaranteed like in a simulation study.

\section{Entropy Balancing}\label{entropy}

Entropy balancing has emerged as a popular method for estimating weights in a variety of contexts, particularly for estimating the average treatment effect of the treated.\cite{hainmueller_entropy_2012,zhao_entropy_2017} Entropy balancing has also previously been introduced for evaluating indirect comparisons of randomized trials, though in this case it is referred to as a method of moment estimator for the inverse odds of sampling when the probability of trial participation follows a a logit model.\cite{signorovitch_comparative_2010} This method of moment estimator just so happens to be the dual solution to an entropy balancing primal problem. We prefer to use the term entropy balancing as it more concisely describes the underlying constrained convex optimization problem that must be solved in order to balance the covariate distribution,
\begin{equation}\label{primal}
\begin{split}
\text{minimize} &\quad \sum_{\{i:S_i = 1\}} \gamma_i\log{\gamma_i} - \gamma_i\\ 
\text{subject to} &\quad \sum_{\{i:S_i = 1\}} \mathbf{A}_{i}\gamma_i = \mathbf{B}
\end{split} 
\end{equation}
Note that the criterion distance in (\ref{primal}) is the entropy function, hence the ``entropy" in entropy balancing. The constraints in (\ref{primal}), represented by the vectors $\mathbf{A}_i$ and $\mathbf{B}$, can be constructed to satisfy some moment balancing conditions, hence the ``balancing" aspect of entropy balancing. For example we can set $\mathbf{A}_{i} = \mathbf{c}(\mathbf{X}_i)$ and $\mathbf{B} = \hat{\boldsymbol{\theta}}_0$ so that the weighted sample moments of $\mathbf{c}(\mathbf{X}_i)$ for all $i \in \{i:S_i = 1\}$ are equal to the sample moments $\mathbf{c}(\mathbf{X}_i)$ for all $i \in \{i:S_i = 0\}$. This specific choice of $\mathbf{A}_i$ and $\mathbf{B}$ is the primal problem for the previously proposed method of moments estimator.\cite{signorovitch_comparative_2010} Entropy balancing and the method of moment estimator for evaluating indirect comparisons are often conflated due to the different Lagrangian dual solutions one can arrive at while solving (\ref{primal}), one of which we will get to later in this section. Nevertheless, due to the strict convexity of the criterion function, the solution to (\ref{primal}) is unique and hence the dual solution must also be unique.\cite{josey_framework_2020} This result was also made explicit specifically when the convex criterion function is the entropy function.\cite{phillippo_equivalence_2020}

The dual solution for the method of moments estimator only requires the target sample moments of the covariates. For this balancing solution, denote $\tilde{\mathbf{c}}(\mathbf{X}_i) = (2Z_i - 1, \mathbf{X}_i)$ and $\tilde{\boldsymbol{\theta}}_0 = \left(0, \hat{\boldsymbol{\theta}}^T_0\right)^T$. The method of moments estimator first solves the Lagrangian dual problem
\begin{equation}\label{MOM}
\hat{\boldsymbol{\lambda}} = \argmax_{\boldsymbol{\lambda} \in \Re^{m + 1}} \ \sum_{\{i:S_i = 1\}}  \left[ -\exp\left(-\tilde{\mathbf{c}}(\mathbf{X}_i)^T\boldsymbol{\lambda}  \right) - \tilde{\boldsymbol{\theta}}^{T}_0 \boldsymbol{\lambda}  \right], 
\end{equation} 
which in turn is used to estimate the sampling weights, $\hat{\gamma}^{\text{MOM}}_i = \exp\left[-\tilde{\mathbf{c}}(\mathbf{X}_i)^T \hat{\boldsymbol{\lambda}}  \right]$ for all $i \in \{i:S_i = 1\}$. We can then use a Horvitz-Thompson type estimator similar to the inverse odds of sampling weights to estimate $\tau_{\text{TATE}}$, \[ \hat{\tau}_{\text{MOM}} = \sum_{\{i:S_i = 1\}} \frac{\hat{\gamma}^{\text{MOM}}_i(2Z_i - 1)Y_i}{\sum_{\{i:S_i = 1\}} \hat{\gamma}^{\text{MOM}}_iZ_i}. \] In Signorovitch \textit{et al}., Assumptions \ref{exchange}-\ref{sita} along with Assumption \ref{odds}, are necessary to establish the consistency of $\hat{\tau}_{\text{MOM}}$ for $\tau_{\text{TATE}}$.\cite{signorovitch_comparative_2010} More recent work can be adapted to show that this estimator is also consistent when Assumption \ref{linear} holds, thus achieving the doubly-robust property.\cite{dong_integrative_2020} 

Our proposed adaptation to this entropy balancing solution instead sets $\mathbf{A}_{i} = \left(\mathbf{A}_{i0}^T, \mathbf{A}_{i1}^T\right)^T$ with $\mathbf{A}_{i0} = (1 - Z_i) \mathbf{c}(\mathbf{X}_i)$ and $\mathbf{A}_{i1} = Z_i\mathbf{c}(\mathbf{X}_i)$ while $\mathbf{B} = \left(\hat{\boldsymbol{\theta}}^T_0, \hat{\boldsymbol{\theta}}^T_{0}\right)^T$ to solve (\ref{primal}) using the following separable Lagrangian dual problem,
\begin{equation}\label{dual}
\begin{split}
\hat{\boldsymbol{\lambda}}_0 &= \argmax_{\boldsymbol{\lambda} \in \Re^{m}} \ \sum_{\{i:S_i = 1\}} \left\{ -\exp\left[-(1 - Z_i) \mathbf{c}(\mathbf{X}_i)^T \boldsymbol{\lambda}\right] - \hat{\boldsymbol{\theta}}^T_0 \boldsymbol{\lambda} \right\} \enskip \text{and} \\
\hat{\boldsymbol{\lambda}}_1 &= \argmax_{\boldsymbol{\lambda} \in \Re^{m}} \ \sum_{\{i:S_i = 1\}} \left[ -\exp\left(-Z_i\mathbf{c}(\mathbf{X}_i)^T \boldsymbol{\lambda}\right) - \hat{\boldsymbol{\theta}}^{T}_0 \boldsymbol{\lambda} \right].
\end{split}
\end{equation}
The empirical sampling weights are subsequently found with
\begin{equation}\label{weights}
\hat{\gamma}^{\text{CAL}}_i = \exp\left[-(1 - Z_i) \mathbf{c}(\mathbf{X}_i)^T \hat{\boldsymbol{\lambda}}_{0} -Z_i\mathbf{c}(\mathbf{X}_i)^T \hat{\boldsymbol{\lambda}}_{1}\right] \ \text{for all} \ i \in \ \{i:S_i = 1\} .
\end{equation}
The estimator for $\tau_{\text{TATE}}$ using these estimated sampling weights is the same Horvitz-Thompson type estimator used by both the MOM and the IOSW approaches,
\begin{equation}\label{HT}
\hat{\tau}_{\text{CAL}} = \sum_{\{i:S_i = 1\}} \frac{\hat{\gamma}^{\text{CAL}}_i(2Z_i - 1)Y_i}{\sum_{\{i:S_i = 1\}}\hat{\gamma}^{\text{CAL}}_iZ_i}. 
\end{equation} 
Notice that the covariate distributions are balanced between treatment groups and between the target sample and the trial participants. This is in contrast with the MOM estimator which only balances the covariate distribution between the target sample and the trial participants. This alteration to $\hat{\tau}_{\text{MOM}}$ remains doubly-robust for estimating $\tau_{\text{TATE}}$ given either Assumption \ref{linear} or \ref{odds}. The double-robustness property of $\hat{\tau}_{\text{CAL}}$ is examined more closely in the Supplementary Material. The alteration to the MOM estimator is also motivated by the equivalence of (\ref{dual})-(\ref{HT}) to the exponential tilting estimator implicitly proposed by Chan \textit{et al}., which is why we refer to it as the calibration (CAL) estimator.\cite{chan_globally_2016} Recall that $\tau_{\text{TATE}}$ is a special case of a weighted average treatment. According to Theorem 3 of Chan \textit{et al}., if we can can uniformly approximate $\rho(\mathbf{X}_i)$, $\mu_1(\mathbf{X}_i)$, and $\mu_0(\mathbf{X}_i)$ using a sufficiently rich basis represented by the balance functions $c_1(\mathbf{X}_i), c_2(\mathbf{X}_i), \ldots, c_m(\mathbf{X}_i)$ (i.e. the number of balancing functions $m$ increases with $n$) while assuming mild conditions about the data generating processes, then the estimate of $\tau_{\text{TATE}}$ with $\hat{\tau}_{\text{CAL}}$ will achieve the efficiency bound in (\ref{semi}).\cite{chan_globally_2016} This efficiency property is not shared by the method of moments version of entropy balancing, further motivating the calibration version.\cite{dong_integrative_2020} Additional details on how to estimate the variance appears in the Supplementary Material.

There are a few reasons why we use the relative entropy over other criterion distance functions for transporting experimental results. The first is due to the resemblance of (\ref{weights}) to the inverse odds of sampling prescribed under Assumption \ref{odds}. This relationship has been noted in several other articles. \cite{signorovitch_comparative_2010, zhao_entropy_2017} Another reason for using the relative entropy is the guarantee that the estimated sampling weights will always be positive. Another suggestion might be to construct a Lagrangian dual using the Euclidean distance as the criterion function to get a different version of (\ref{primal}) and (\ref{dual}). However, the support for the Euclidean distance is the real numbers, implying that negative weights are feasible in such a setup. Adding the necessary constraint that $\gamma_i > 0$ for all $i = 1,2,\ldots,n$ makes the optimization problems in (\ref{primal}) and (\ref{dual}) less straightforward to solve.  

Now consider the setting where we are provided only the sample covariate moments from the target sample. Assuming that $\hat{\boldsymbol{\theta}}_0$ is fixed results in an inflated Type I error rate for inferences of $\tau_{\text{TATE}}$. The one exception to this rule is when $\hat{\boldsymbol{\theta}}_0 = \boldsymbol{\theta}_0$ with zero variability. In other words, we would need to estimate $\hat{\boldsymbol{\theta}}_0$ over the entire target population. If we are provided individual-level covariate data from the target sample, then we may derive a variance estimator for estimates of $\tau_{\text{TATE}}$ as opposed to $\tau_{\text{TSATE}}$. Despite this shortfall, the estimators (\ref{dual}) - (\ref{HT}) remain consistent for $\tau_{\text{TATE}}$ in either setting. The same rule applies for both the OM approach and the MOM estimator since neither of these methods necessarily require the complete individual-level covariate data. A more concrete demonstration of this phenomenon is shown in Section \ref{sim_2}.

\section{Numerical Examples}\label{simulation}

\subsection{Simulation Study}\label{sim_1}

In this section we present a simulation study to better understand the performance of entropy balancing techniques compared with the alternative methods illustrated in Section \ref{previous}. We consider four experimental scenarios that test the consistency and efficiency of the estimators on finite-samples by altering the data generating processes. \text{red}{These scenarios all make the conditional constancy of absolute effects assumption defined in (\ref{absolute}) to ensure compatibility between all the methods in consideration.}

The first scenario establishes a baseline. For $i = 1,2,\ldots,n$, let $(X_{i0}|S_i = 0) \sim \mathcal{N}(-1, 4)$,  $(X_{i1}|S_i = 0) \sim \text{Bin}(1, 0.6)$, $(X_{i2}|S_i = 0) \sim \mathcal{N}(0, 1)$, and $(X_{i3}|S_i = 0) \sim \text{Bin}(1, 0.5)$. Let $(X_{i0}|S_i = 1) \sim \mathcal{N}(1, 4)$,  $(X_{i1}|S_i = 1) \sim \text{Bin}(1, 0.4)$, $(X_{i2}|S_i = 1) \sim \mathcal{N}(0, 1)$, and $(X_{i3}|S_i = 1) \sim \text{Bin}(1, 0.5)$. We generate the treatment assignment by sampling $Z_i \sim \text{Bin}(1, 0.5)$. The conditional mean of the potential outcomes are constructed as
\begin{equation}\label{mu}
\begin{split}
\mu_0(\mathbf{X}_i) &= 10 - 3X_{i0} - X_{i1} + X_{i2} + 3X_{i3} \enskip \text{and} \\
\mu_1(\mathbf{X}_i) &= \mu_0(\mathbf{X}_i) + 5 + 3X_{i0} - X_{i1} + X_{i2} - 3X_{i3}.
\end{split}
\end{equation}
Gaussian potential outcomes for each experimental scenario are generated by sampling $Y_{i}(0) \sim \mathcal{N}[\mu_0(\mathbf{X}_i), \sigma^2]$ and $Y_{i}(1) \sim \mathcal{N}[\mu_1(\mathbf{X}_i), \sigma^2]$, with the observed outcome determined by $Y_i = Z_iY_i(1) + (1 - Z_i)Y_i(0)$ for each $i = 1,2,\ldots,n$. We discard the $n_0$ values of $Y_i$ and $Z_i$ for all $i \in \{i:S_i = 0\}$. We will refer to this set of conditions with the label ``baseline". Unless stated otherwise, $S_i$, $\mathbf{X}_i = (X_{i0}, X_{i1}, X_{i2}, X_{i3})$, $Y_i$, and $Z_i$ are provided to estimate $\hat{\gamma}_i^{\text{PS}}$, $\hat{\gamma}_i^{\text{MOM}}$, $\hat{\gamma}_i^{\text{CAL}}$, $\hat{\mu}_1(\mathbf{X}_i)$, $\hat{\mu}_0(\mathbf{X}_i)$, $\tilde{\mu}_1(\mathbf{X}_i)$, and $\tilde{\mu}_0(\mathbf{X}_i)$ which are required in the estimators described in \ref{previous} and \ref{entropy}.

In the scenarios labeled ``positivity", we increase the difference between the two covariate distributions by substituting $(X_{i0}|S_i = 0) \sim \mathcal{N}(1, 2)$, $(X_{i0}|S_i = 1) \sim \mathcal{N}(-1, 1)$, $(X_{i1}|S_i = 0) \sim \text{Bin}(1, 0.2)$, and $(X_{i1}|S_i = 1) \sim \text{Bin}(1, 0.8)$ for the respective covariates into the data generating mechanisms. This alteration will test the sensitivity of each method on the intrinsic limitations associated with Assumption \ref{positivity}. For the scenario labeled ``sparse", we provide each method an additional set of covariates that do not affect the responses. The potential outcomes are still determined from (\ref{mu}) with the original covariate values, yet the different estimators must also accommodate the additional noise covariates of $(X_{ir}|S_i = 0) \sim (X_{i(r-4)}|S_i = 1)$ and $(X_{ir}|S_i = 1) \sim (X_{i(r-4)}|S_i = 0)$ for $r \in \{4,5,6,7\}$. Each of the estimators are provided data for $(X_{i0}, X_{i1}, \ldots, X_{i7})$ in addition to $Y_i$ $Z_i$, and $S_i$. In the scenarios labeled ``missing", we generate the outcomes according to (\ref{mu}) yet we provide each method only $(X_{i0}, X_{i2})$ and omit $X_{i0}$ and $X_{i2}$  while estimating $\tau_{\text{TATE}}$. Note that this means we omit one of the effect modifiers, $X_{i1}$. Next, we formulate scenarios which misspecify the outcome model (``outcome"). To do so, we generate outcomes according to the model
\begin{equation}\label{mu_int}
\begin{split}
\mu_0(\mathbf{X}_i) &=  10 - 3U_{i0} - X_{i1} + U_{i2} + 3X_{i3} \enskip \text{and} \\
\mu_1(\mathbf{X}_i) &= \mu_0(\mathbf{X}_i) + 5 + 3U_{i0} - X_{i1} + U_{i2} - 3X_{i3}
\end{split}
\end{equation}
where $U_{i0} = \exp(-X_{i0}/4 + X_{i2}/4)$ and $U_{i2} = (X_{i0}/2 - X_{i2}/2)^2$. Both $U_{i0}$ and $U_{i2}$ are standardized across both samples to have a mean of 0 and variance 1 in the combined trial and target samples. We then provide each method the original covariate values $(X_{i0}, X_{i1},X_{i2},X_{i3})$ for all $i = 1,2,\ldots,n$. On the other hand, in the sampling misspecification scenario (``sampling"),  we provide each method data for $(U_{i0}, X_{i1}, U_{i2}, X_{i3})$ while the outcomes are still generated by the model in (\ref{mu_int}). The standardization step is key to ensure that the true magnitude of the differences between the sample covariate distributions are never fully expressed by the sampling model. In addition to varying the scenarios that test the violations to the assumptions in Section \ref{assumptions}, we also vary $n_0 \in \{500,1000\}$ and $n_1 \in \{500,1000\}$, creating $24$ different conditions for which we will generate $1000$ replications.

We report the average bias and empirical mean squared error of the average treatment effect estimates across the $1000$ iterations for each scenario. The average model and empirical standard errors are provided in tables S1 and S2 in the Supplementary Materials. The model standard errors are obtained using a sandwich variance estimator for each estimate in every iteration of the simulation. The empirical standard errors are the standard deviations of the estimates from each estimator pooled across the iterations of a given scenario. The five methods we compare for estimating the target average treatment effect are: Inverse Odds of Sampling Weights (IOSW), G-Computation (OM), Augmented Inverse Odds of Sampling Weights (DR), Targeted Maximum Likelihood Estimation (TMLE), Method of Moments (MOM), and Calibration (CAL). Additionally, for IOSW, OM, DR, and TMLE, both $\hat{\mu}_1(\mathbf{X}_i)$ and $\hat{\mu}_0(\mathbf{X}_i)$ are fit by regressing $Y_i$ onto the covariates provided in each scenario with data from $S_i = 1$ and stratified by the $Z_i$. $\hat{\rho}(\mathbf{X}_i)$ is fit with logistic regression using covariates that predict $S_i$. The standard errors are estimated using robust sandwich variance estimators. For the non-entropy balancing type estimators, \texttt{R} code for finding variance estimates are provided in the existing literature.\cite{rudolph_robust_2017, dahabreh_extending_2020}

\begin{table}
\centering
\scriptsize
\begin{tabular}{cccccccccc}
\hline
$n_0$ & $n_1$ & Scenario & $\tau_{\text{TATE}}$ & IOSW & OM & DR & TMLE & MOM & CAL \\ \hline
500 & 1000 & baseline & -0.1 & 0.03 (0.47) & 0.00 (0.09) & 0.00 (0.10) & 0.00 (0.10) & 0.00 (0.33) & 0.00 (0.10) \\
500 & 1000 & missing & -0.1 & 0.23 (0.48) & 0.19 (0.16) & 0.19 (0.17) & 0.19 (0.17) & 0.19 (0.36) & 0.19 (0.17) \\
500 & 1000 & outcome & 8.9 & -0.03 (0.41) & -1.44 (2.23) & -0.08 (0.72) & -0.03 (0.26) & -0.05 (0.24) & -0.08 (0.20) \\
500 & 1000 & positivity & -0.3 & 0.27 (1.51) & -0.01 (0.08) & -0.01 (0.19) & -0.01 (0.38) & 0.00 (0.81) & -0.01 (0.19) \\
500 & 1000 & sample & 4.5 & 0.64 (3.61) & 0.00 (0.03) & 0.00 (0.05) & -0.02 (0.59) & 0.00 (0.06) & 0.00 (0.04) \\
500 & 1000 & sparse & -0.1 & 0.08 (1.18) & -0.01 (0.11) & -0.01 (0.16) & -0.02 (0.20) & -0.03 (0.63) & -0.01 (0.16) \\
1000 & 200 & baseline & -0.1 & 0.15 (2.04) & 0.00 (0.14) & 0.00 (0.18) & -0.01 (0.18) & -0.02 (1.18) & -0.01 (0.18) \\
1000 & 200 & missing & -0.1 & 0.30 (1.94) & 0.18 (0.27) & 0.20 (0.33) & 0.20 (0.33) & 0.19 (1.07) & 0.20 (0.33) \\
1000 & 200 & outcome & 8.9 & -0.10 (1.66) & -1.45 (2.87) & -0.22 (2.87) & -0.15 (1.14) & -0.15 (0.87) & -0.26 (0.76) \\
1000 & 200 & positivity & -0.3 & 0.91 (4.80) & 0.03 (0.24) & 0.05 (0.49) & 0.05 (3.10) & 0.11 (3.17) & 0.08 (0.62) \\
1000 & 200 & sample & 3.2 & 0.37 (2.63) & -0.01 (0.10) & -0.01 (0.14) & 0.07 (1.57) & 0.00 (0.20) & -0.01 (0.13) \\
1000 & 200 & sparse & -0.1 & 0.31 (4.14) & -0.01 (0.20) & -0.01 (0.37) & -0.01 (0.84) & -0.05 (2.67) & 0.00 (0.41) \\
1000 & 1000 & baseline & -0.1 & 0.01 (0.48) & 0.00 (0.06) & 0.00 (0.07) & 0.00 (0.07) & -0.02 (0.29) & 0.00 (0.07) \\
1000 & 1000 & missing & -0.1 & 0.21 (0.46) & 0.19 (0.11) & 0.19 (0.13) & 0.19 (0.13) & 0.19 (0.30) & 0.19 (0.13) \\
1000 & 1000 & outcome & 8.9 & -0.03 (0.43) & -1.41 (2.14) & -0.06 (0.75) & -0.03 (0.23) & -0.05 (0.23) & -0.08 (0.20) \\
1000 & 1000 & positivity & -0.3 & 0.19 (1.74) & 0.00 (0.06) & 0.00 (0.16) & -0.01 (0.31) & -0.01 (0.81) & 0.00 (0.16) \\
1000 & 1000 & sample & 4.0 & 0.78 (4.22) & 0.00 (0.02) & -0.01 (0.04) & -0.12 (1.89) & -0.01 (0.04) & -0.01 (0.03) \\
1000 & 1000 & sparse & -0.1 & 0.12 (1.03) & -0.01 (0.07) & -0.01 (0.12) & -0.01 (0.14) & 0.00 (0.57) & -0.01 (0.11) \\ \hline
\end{tabular}
\caption{Average bias and (mean squared error) of $\tau_{\text{TATE}}$ estimates. The scenarios where $n_1 = 200$ and $n_0 = 500$ are detailed in Figure \ref{ATE-plot}.}\label{table-1}
\end{table}

\begin{figure}
	\centering
	\includegraphics[width=1\linewidth]{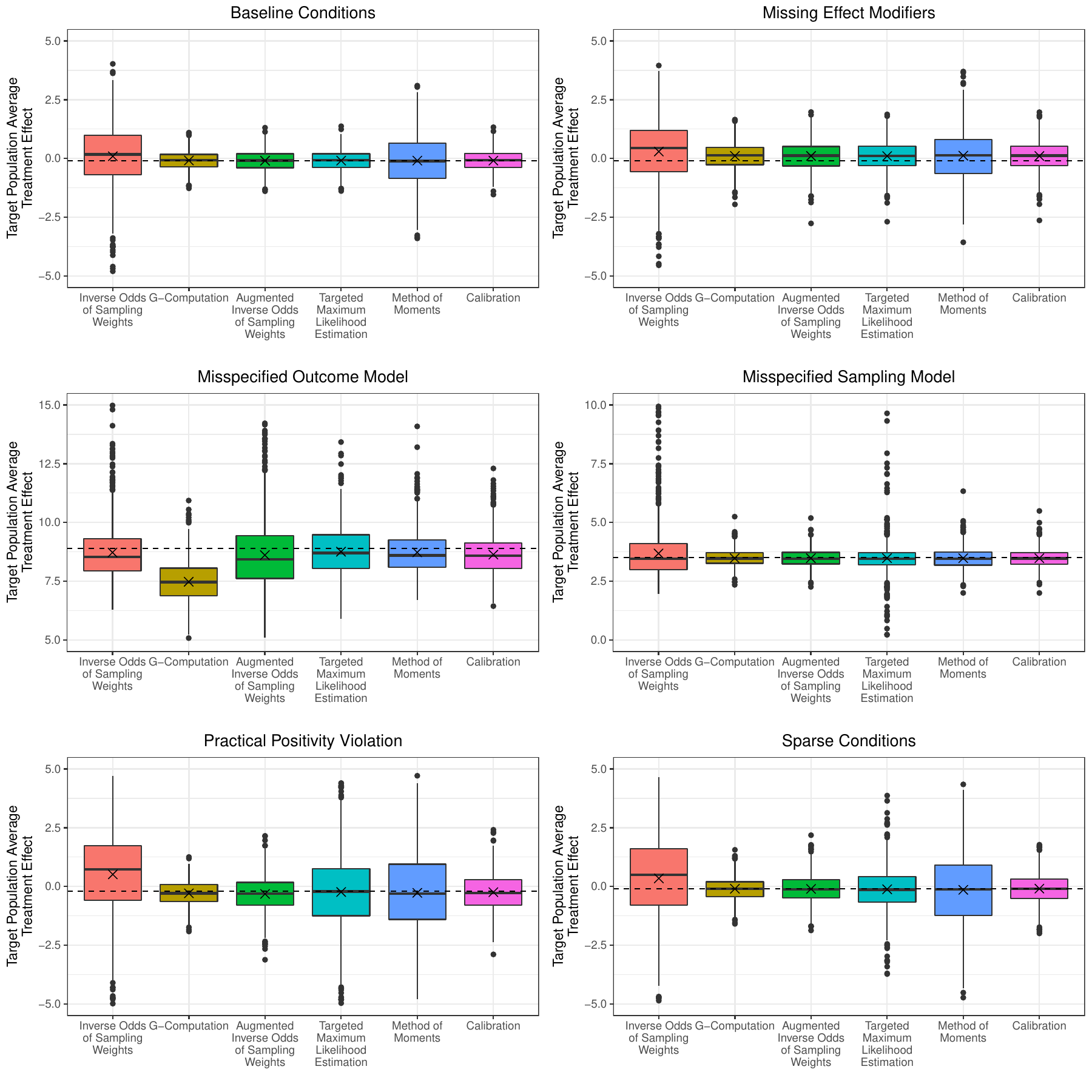}
	\caption{Estimates of the target population average treatment effects over the 1000 iterations of the simulation study described in Section \ref{sim_1}. The dashed line demarcates the true target population average treatment effect for each scenario while the x is the average of the estimates. These estimates are drawn from cases when $n_1 = 200$ and $n_0 = 500$.}\label{ATE-plot}
\end{figure}

The average bias and mean squared errors of the experiment are summarized in Table \ref{table-1}. A visual comparison for a subset of the results featured in Table \ref{table-1} where $n_1 = 200$ and $n_0 = 500$ appear in the boxplots of Figure \ref{ATE-plot}. Each method produces consistent estimates under the baseline scenario. However, each method also has its short-comings. First, we can see that IOSW produce highly variable estimates in cases where the positivity assumption (Assumption \ref{positivity}) is practically violated and is biased in cases when the sampling scenario is misspecified. On the other hand, the OM approach is biased when the outcome model is misspecified. TMLE, DR, MOM, and CAL all appear to produce unbiased estimates of the average treatment effect in every scenario. This is interesting for the MOM estimator since this would imply that it is also doubly-robust in terms of consistency. Some insight into why this might be is provided elsewhere.\cite{dong_integrative_2020} However, we can see in Table \ref{table-1} that CAL had either the same or smaller mean squared errors over TMLE and MOM. In some scenarios DR did have smaller errors than CAL. Nevertheless,  The OM approach had the smallest mean squared errors across most scenarios, other than in the scenarios where we misspecify the outcome model naturally. When the models miss (or ignore) some of the effect modifiers, we see that every method we test produces biased estimates of the target population average treatment effect. This particular scenario emphasizes the results of these estimators when both the outcome and the sampling models are misspecified, even when missing a single effect modifier.

There is a downside to the so-called calibration version of entropy balancing. In the sparse and positivity violation scenarios, the number of models that converge decreases considerably. When $n_1 = 200$ and $n_0 = 500$, CAL was only able to find a solution in 64.2\% of the iterations under the sparse scenario and 49.0\% of the iterations when positivity is practically violated. When $n_1 = 200$ and $n_0 = 1000$ we observe a 66.0\% and 48.1\% rate of convergence in the sparse and positivity scenarios, respectively. Otherwise, the calibration approach to entropy balancing converged in each iteration for every other scenario. Meanwhile, the method of moments approach to entropy balancing also failed to converge in approximately 7.5\% of the scenarios in both of the positivity scenarios where the calibration estimator often failed to converge.

\subsection{Coverage Probabilities of $\tau_{\text{TATE}}$ and $\tau_{\text{TSATE}}$}\label{sim_2}

Consider the baseline scenario in the previous set of simulations. Using the individual-level data from the trial sample, and the target sample covariate moments, we demonstrate how inferences for $\tau_{\text{TATE}}$ will have an inflated Type I error. We do so by finding the empirical coverage probability of both $\tau_{\text{TSATE}}$ and $\tau_{\text{TATE}}$ with both of the entropy balancing approaches described in Section \ref{entropy}. Robust sandwich variance estimators are used to construct the confidence intervals. The coverage probability is obtained by averaging over the indicator variable generated by whether the resulting confidence interval about the average treatment effect estimate covers either of $\tau_{\text{TSATE}}$ or $\tau_{\text{TATE}}$ at each iteration. This will demonstrate why entropy balancing can only be used to infer upon the target sample average treatment effect instead of the target population average treatment effect unless the entire individual-level data about the balance functions in both the target and trial samples is available. For this simulation experiment, we set the target and trial sample sizes at $n_0 \in \{500,1000,10000\}$ and $n_1 \in \{1000,10000\}$, respectively. We use large sample sizes to ensure the accuracy of the robust variance estimator.

\begin{table}
\centering
\scriptsize
\begin{tabular}{cccccccccc}
\hline
\multirow{2}{*}{$n_0$} & \multirow{2}{*}{$n_1$} &  & \multicolumn{4}{c}{Without Individual Level Data} &  & \multicolumn{2}{c}{With Individual Level Data} \\ \cline{4-7} \cline{9-10} 
 &  &  & MOM $\tau_{\text{TSATE}}$ & MOM $\tau_{\text{TATE}}$ & CAL $\tau_{\text{TSATE}}$ & CAL $\tau_{\text{TATE}}$ &  & MOM $\tau_{\text{TATE}}$ & CAL $\tau_{\text{TATE}}$ \\ \hline
500 & 1000 &  & 0.926 & 0.882 & 0.929 & 0.623 &  & 0.937 & 0.943 \\
500 & 10000 &  & 0.956 & 0.667 & 0.936 & 0.289 &  & 0.948 & 0.960 \\
1000 & 1000 &  & 0.922 & 0.897 & 0.938 & 0.761 &  & 0.928 & 0.951 \\
1000 & 10000 &  & 0.929 & 0.770 & 0.938 & 0.372 &  & 0.945 & 0.942 \\
10000 & 1000 &  & 0.928 & 0.923 & 0.930 & 0.902 &  & 0.923 & 0.929 \\
10000 & 10000 &  & 0.930 & 0.912 & 0.960 & 0.791 &  & 0.935 & 0.958 \\ \hline
\end{tabular}
\caption{Coverage Probabilities of $\tau_{\text{TSATE}}$ and $\tau_{\text{TATE}}$ using Entropy Balancing and Outcome Modeling Techniques.}\label{table-2}
\end{table}

The results in Table \ref{table-2} show how modifying $n_1$ and $n_0$ affects the coverage probabilities for $\tau_{\text{TSATE}}$ and $\tau_{\text{TATE}}$ for the setting where we are given the target sample covariate moments. Observe that the coverage probability of $\tau_{\text{TSATE}}$ is dependent on $n_1$ alone - as $n_1$ increases, the coverage probabilities increase. The coverage probability of $\tau_{\text{TATE}}$, on the other hand, appears to be dependent on the ratio between $n_0$ and $n_1$. For inference on $\tau_{\text{TATE}}$, we see the best results occur when $n_1$ is small relative to $n_0$. When $n_1 = 1000$ and $n_0 = 10000$, the variation of $\hat{\boldsymbol{\theta}}$ has less impact on the total variance, producing the best results. In contrast, when $n_1 = 10000$ and $n_0 = 10000$, the variation of $\hat{\boldsymbol{\theta}}$ has a greater impact, resulting in a decreased probability of coverage. This observation is only compounded in cases where $n_1 > n_0$. This leads us to believe that $n_1$ needs to be sufficiently large while also remaining small compared to $n_0$ in order to be effective for inferring on $\tau_{\text{TATE}}$. When we adjust the sandwich estimator to incorporate individual-level covariate data from the target sample, we see that the accuracy of coverage probability is now tied to the total sample size $n = n_0 + n_1$, which is typical for robust variance estimators as they are derived under asymptotic conditions.

\section{Transporting Results of ACCORD-BP study to the US Population}\label{illustrative}

Translating clinical trial results to clinical care is particularly challenging when the results of two studies conducted for similar indications and treatments yield conflicting conclusions. For example, the optimal approach to hypertension treatment remains unclear, partly due to conflicting clinical trial results. The Systolic Blood Pressure Intervention Trial (SPRINT) and the Action to Control Cardiovascular Risk in Diabetes Blood Pressure (ACCORD-BP) trial both randomized participants with hypertension to intensive ($<120$ mmHg) or conventional ($<140$ mmHg) blood pressure control targets. The study populations differed in that ACCORD-BP was limited to diabetes patients while SPRINT excluded diabetes patients. The two similarly designed studies in differing populations had different results: SPRINT, but not ACCORD-BP, found an association of intensive blood pressure control with several clinically meaningful outcomes including cardiovascular disease events.\cite{the_accord_study_group_effects_2010, the_sprint_study_group_randomized_2015} Importantly, the ACCORD-BP trial was enriched for individuals at high cardiovascular disease risk aside from the presence of diabetes, raising the question of whether the result of the trial applies to a more general diabetes patient population. Thus, transporting the ACCORD-BP trial to the general US population of diabetes patients may provide insight into hypertension management for individuals with diabetes and help reconcile the discrepant trial results.

To address this question, Berkowitz \textit{et al}. used inverse odds of sampling weights (IOSW) to transport the ACCORD-BP trial to a sample of US diabetes patients drawn from the US National Health and Nutrition Examination Survey (NHANES).\cite{berkowitz_generalizing_2018} They found that weighting the ACCORD-BP sample to reflect the diabetes patient sample in NHANES yielded intervention effects more similar to those observed in the SPRINT trial of non-diabetes patients than in the unweighted ACCORD-BP trial. We use this previously demonstrated application of transportability methods to ACCORD-BP as a real-world application of the entropy balancing (CAL) methods described here. In our applied example, we transport four-year post randomization risk difference estimates of total mortality observed in the ACCORD-BP trial\cite{the_accord_study_group_effects_2010} to a sample of US diabetes patients drawn from the NHANES cohort.\cite{berkowitz_generalizing_2018} We use two methods for transporting the results of ACCORD-BP to NHANES - IOSW and entropy balancing (CAL). Furthermore, using entropy balancing we provide confidence intervals about the target sample average treatment effect and the target population average treatment effect. Recall that the former estimand does not require any individual-level data from the NHANES sample.

\begin{table}
\centering
\scriptsize
\begin{tabular}{lcccc}
\hline
Variables & NHANES & ACCORD-BP & IOSW ACCORD-BP & CAL ACCORD-BP \\ \hline
Baseline age, yrs & $59.65 \pm 13.70$ & $62.84 \pm 6.74$ & $61.50 \pm 6.66$ & $59.61 \pm 6.91$ \\
Female & $48.9$ & $47.1$ & $48.0$ & $48.9$ \\
Race/Ethnicity &  &  &  &  \\
\multicolumn{1}{r}{Non-Hispanic white} & $62.6$ & $58.7$ & $51.6$ & $62.5$ \\
\multicolumn{1}{r}{Non-Hispanic black} & $15.2$ & $24.0$ & $19.6$ & $15.2$ \\
\multicolumn{1}{r}{Hispanic} & $15.2$ & $6.8$ & $22.3$ & $15.2$ \\
\multicolumn{1}{r}{Asian/multi/other} & $7.0$ & $10.5$ & $6.5$ & $7.1$ \\
Insurance & $86.8$ & $85.0$ & $84.6$ & $86.6$ \\
Smoking status &  &  &  &  \\
\multicolumn{1}{r}{Never} & $51.4$ & $44.6$ & $52.9$ & $51.4$ \\
\multicolumn{1}{r}{Former} & $33.1$ & $42.6$ & $30.9$ & $33.1$ \\
\multicolumn{1}{r}{Current} & $15.5$ & $12.8$ & $16.2$ & $15.5$ \\
Education &  &  &  &  \\
\multicolumn{1}{r}{Less than HS} & $25.7$ & $16.3$ & $30.9$ & $25.7$ \\
\multicolumn{1}{r}{HS diploma} & $27.1$ & $27.0$ & $26.4$ & $27.1$ \\
\multicolumn{1}{r}{Some college} & $29.3$ & $32.4$ & $26.5$ & $29.3$ \\
\multicolumn{1}{r}{College diploma or higher} & $17.9$ & $24.3$ & $16.2$ & $17.9$ \\
History of CHF & $7.7$ & $4.2$ & $11.0$ & $7.7$ \\
History of MI & $10.5$ & $13.6$ & $11.4$ & $10.5$ \\
History of stroke & $7.9$ & $6.4$ & $7.5$ & $7.8$ \\
Years with diabetes & $7.49 \pm 9.20$ & $10.88 \pm 7.83$ & $10.05 \pm 7.26$ & $7.50 \pm 6.51$ \\
BMI, kg/$\text{m}^2$ & $32.80 \pm 7.31$ & $32.10 \pm 5.47$ & $32.07 \pm 5.52$ & $32.80 \pm 5.78$ \\
SBP, mm Hg & $130.05 \pm 19.15$ & $139.33 \pm 15.61$ & $133.94 \pm 14.42$ & $129.67 \pm 13.98$ \\
DBP, mm Hg & $69.50 \pm 12.96$ & $75.86 \pm 10.28$ & $71.53 \pm 9.57$ & $69.44 \pm 9.55$ \\
HDL, mg/dl & $49.11 \pm 13.46$ & $46. 049 \pm 13.68$ & $51.60 \pm 17.24$ & $49.08 \pm 17.72$ \\
LDL, mg/dl & $103.83 \pm 36.03$ & $110.70 \pm 36.52$ & $105.42 \pm 31.33$ & $103.59 \pm 33.31$ \\
Triglycerides, mg/dl & $148.93 \pm 76.13$ & $193.36 \pm 174.21$ & $125.40 \pm 68.01$ & $147.21 \pm 95.52$ \\
FPG, mg/dl & $151.88 \pm 54.62$ & $174.81 \pm 57.66$ & $171.54 \pm 57.47$ & $151.11 \pm 47.30$ \\
$\text{HbA}_{1\text{c}}$, \% & $7.16 \pm 1.64$ & $8.34 \pm 1.09$ & $7.94 \pm 0.95$ & $7.16 \pm 0.75$ \\
Estimated GFR, ml/min & $87.46 \pm 28.11$ & $91.64 \pm 29.83$ & $84.99 \pm 21.16$ & $87.31 \pm 22.66$ \\
Urine albumin to creatinine ratio & $75.44 \pm 481.68$ & $93.84 \pm 333.81$ & $105.89 \pm 427.60$ & $45.32 \pm 204.57$ \\ \hline
\end{tabular}
\caption{Values are mean $\pm$ SD or \%. Means and percentages for NHANES are nationally representative using NHANES sampling weights.}\label{table-3}
\end{table}

\begin{figure}
	\centering
	\includegraphics[scale = 1]{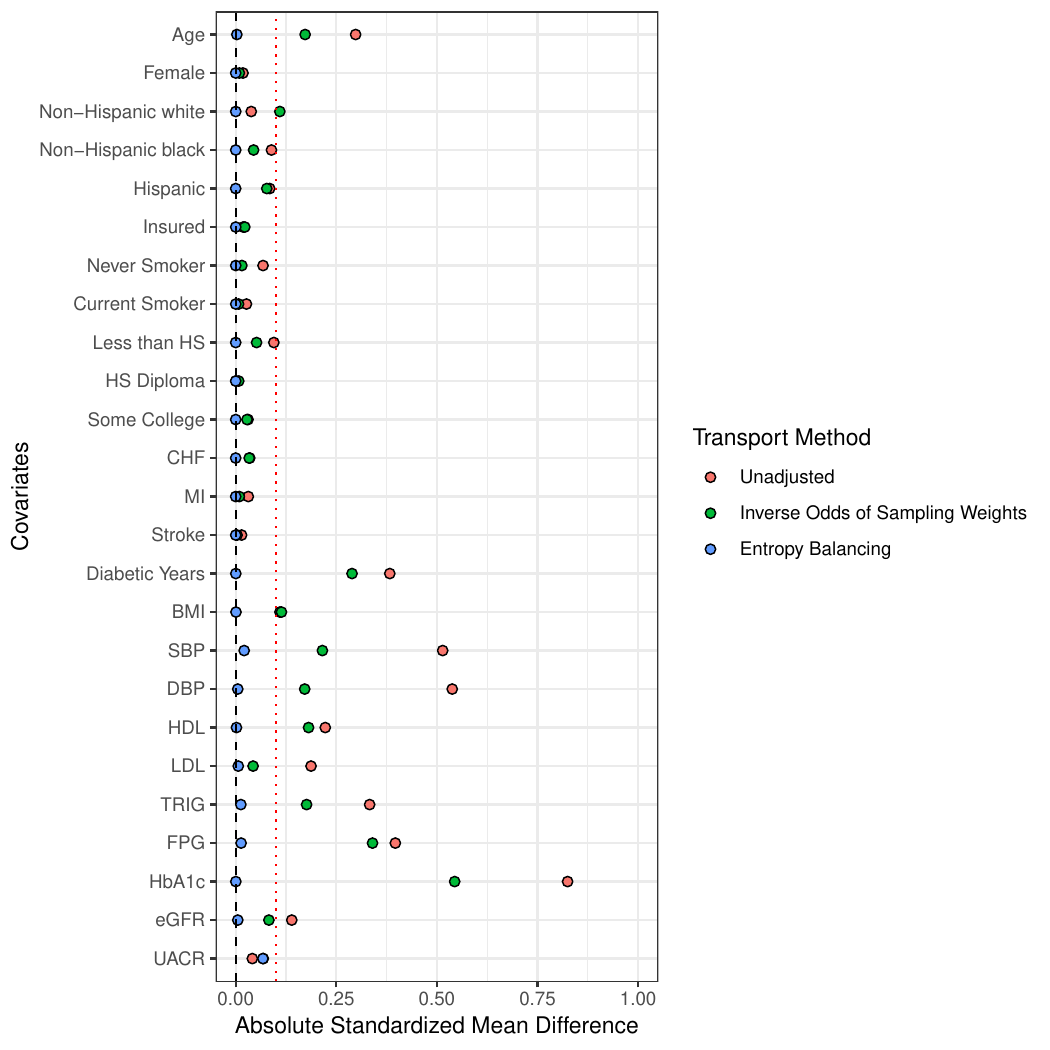}
	\caption{Absolute standardized mean differences for various weighting estimators between NHANES and ACCORD. The red dotted line demarcates an absolute standardized mean difference of 0.1.}\label{bal-plot}
\end{figure}

Table \ref{table-3} shows covariates (which we set as the balance functions) balanced between ACCORD-BP and NHANES, their unweighted sample covariate moments from the NHANES and from the ACCORD-BP data, and the subsequent weighted covariate sample moments of the ACCORD-BP sample after balancing. Compared to ACCORD-BP, the NHANES diabetes sample was younger, more likely to be Hispanic and less likely to be black, more likely to be never smokers, less likely to have a history of myocardial infarction (MI) but more likely to have a history of congestive heart failure (CHF), and had a shorter duration of diabetes and better glycemic control (indicated by hemoglobin A1c) (Table 3). Many of the differences in covariate distributions reflect that ACCORD trial eligibility criteria focused on those with relatively long duration of diabetes and high prevalence of cardiovascular risk factors. Of note, the intensive blood pressure control intervention had a smaller benefit in individuals with pre-existing cardiovascular disease in the SPRINT trial, making it plausible that differences between the ACCORD-BP population and a general population of diabetes patients might modify the effect of the blood pressure intervention.\cite{the_sprint_study_group_randomized_2015} In another study using data from NHANES, hemogloblin A1c was associated with increased risk of all-cause and cause-specific mortality.\cite{palta_hemoglobin_2017} Zoungas \textit{et al.}\cite{zoungas_impact_2014} show that diabetes duration is associated with death while McEwen \textit{et al}.\cite{mcewen_predictors_2012} identified multiple predictors of total mortality such as race, age, and previous cardiovascular events among diabetic patients. These previous findings imply that numerous factors might have the potential for confounding the relationship between sampling and the outcome. The differences in baseline covariates between ACCORD-BP and NHANES are reduced after balancing with both CAL and IOSW. However, the covariate sample moments after CAL weighting consistently matched the NHANES sample more closely than after IOSW weighting (Table \ref{table-3}, Figure \ref{bal-plot}). Small residual differences remain between NHANES and the weighted ACCORD-BP sample, for example with triglycerides and high density lipoproteins (Figure \ref{bal-plot}).

The ACCORD-BP study originally found an increase in four-year mortality of 0.59\% [95\% CI:(-0.75\%, 1.93\%)] in the intensive treatment group. After weighting the ACCORD-BP responses with inverse odds of sampling weights estimated with maximum likelihood, the estimated risk difference on the NHANES population is -1.35\% [95\% CI: (-3.5\%, 0.8\%)]. Using CAL, we observe a risk difference of -0.04\% [95\% CI: (-1.80\%, 1.71\%)] where the confidence interval corresponds to the NHANES sample average treatment effect. The 95\% confidence interval for the NHANES population average treatment effect is (-1.94\%, 1.86\%) when using the individual level covariate data from the NHANES sample. 

Though the total mortality is insignificant at a $0.05$ level of significance, regardless of method, we see changes in the risk difference estimate. The original analysis found an increase in mortality among the intensively treated patients. IOSW weights yielded a decreased total mortality among intensively treated patients in the NHANES population, while CAL weights yielded a nearly null result. These differences seem to indicate the presence of effect modifiers contributing to the effect of blood pressure treatment intensity on mortality. Notice also that the population level estimate is the same as the sample level estimate using entropy balancing. However, the width of the confidence interval is wider for the population level estimate. Nevertheless, the population level estimate from the CAL approach is still narrower than the estimated confidence interval produced by the IOSW approach, indicating an increase in efficiency.

\section{Discussion}\label{discussion}

In this article we have described a doubly-robust method for transporting experimental results borrowed from the entropy optimization literature. We also borrow results from the indirect comparison literature, which allows us to relax the conditional constancy of absolute effects assumption typically applied in the transportability literature and focus our efforts on modeling relative effects with effect modifiers rather than the absolute effects which require both the effect modifiers and any prognostic variables.\cite{rudolph_robust_2017, dahabreh_extending_2020} However, if the sampling model is incorrect, then we would need the conditional constancy of absolute effects assumption to hold in order to get consistent estimates given the doubly-robust property. As a result more emphasis should be placed on correctly specifying the sampling model over the outcome model if modelling choices begin to deviate. The entropy balancing methodology may operate in two settings - when we are presented with the complete individual-level data of the trial sample and either the individual-level covariate data or the covariate sample moments of the target sample. The distinction between the two settings amounts to inferring upon the target population average treatment effect versus the target sample average treatment effect. We showed entropy balancing to be an efficient causal effect estimator in finite-samples through simulation. We also compared two methods for transporting the ACCORD-BP study to the NHANES population. These numerical examples demonstrate some of the practical implications of our work.

The drawback to using entropy balancing for transportability is with the algorithm's rate of convergence. In small samples, the probability that a feasible weighting solution exists decreases, particularly when the positivity assumption is practically violated. One solution applied to covariate balance problems is to use inequality constraints to mitigate treatment group heterogeneity.\cite{wang_minimal_2020} This solution is most useful in high-dimensional settings. There may also be a way to incorporate the method of moment balancing weights into the TMLE framework by substituting $\hat{\boldsymbol{\gamma}}^{\text{MOM}}$ for $\hat{\boldsymbol{\gamma}}^{\text{PS}}$ in (\ref{tmle}). This could eventually set up a targeted maximum likelihood-type estimator that can operate in the setting where we do not have any individual-level data from the target population. In the case where convergence failure occurs due to a high dimension of potential effect modifiers relative to the trial sample size, one should carefully consider balance diagnostics between trial participants and non-participants in much the same way as one identifies potential confounders in an observational study.\cite{brookhart_variable_2006, westreich_role_2011} Additional sensitivity analyses should be employed intermittently to ensure that every effect modifier is accounted for.\cite{nguyen_sensitivity_2018} In the case of positivity violations, it seems that methods that employ more direct implementations of an outcome model, such as the G-compuation and DR approaches, fair better given their ability to extrapolate over the covariate space. Violations of Assumptions \ref{exchange} and \ref{sita} pose a more difficult challenge to evaluate as these assumptions are untestable. Expert level knowledge of the domain area are necessary to ensure that these assumptions will hold with the preferred transportability model.

Future work will address two additional data settings not evaluated here. First, the setting where the target sample contains data from a second randomized experiment, including both the individual-level outcome and the treatment assignment. The process of combining experiments, termed as data-fusion, is beyond what we discuss in this paper but is nevertheless an important problem which we would like to approach with entropy balancing in future research. A second direction for future work is to examine methods for transportability between two observational samples, rather than assuming availability of randomized clinical trial data for the trial sample.\cite{josey2020calibration} In this situation, we would also need to model the probability of treatment within the the observational study representing the ``trial'' sample. We might also seek to relax Assumptions \ref{linear} and \ref{odds} using a nonparametric setup to the problem similar to the sieve approach but instead applied to transportability.\cite{hirano_efficient_2003, chan_globally_2016} Finally, while the average treatment effect estimands under consideration in this manuscript are applicable to various outcomes, including a binary one, more work is needed to generalize many of these estimators to accommodate a non-linear link function for the outcome model.

In summary, entropy balancing provides an approach to transportability that is flexible regarding the applicable data settings and exhibits double robustness in specific scenarios. In particular, entropy balancing yields more precise effect estimates across a range of simulation scenarios when the target population is large than alternative methods using only covariate sample moments from the target population.

\section*{Acknowledgments}

\noindent\textbf{Funding information:} This research was supported in part by the US Department of Veterans Affairs Award IK2-CX001907-01. S.A. Berkowitz's role in the research reported in this publication was supported by the National Institute Of Diabetes And Digestive And Kidney Diseases of the National Institutes of Health under Award Number K23DK109200. D. Ghosh's role reported in the publication was supported by NSF DMS-1914937.

\vspace{1ex}

\noindent\textbf{Disclaimer:} This manuscript will be submitted to the Department of Biostatistics and Informatics in the Colorado School of Public Health, University of Colorado Anschutz Medical Campus, in partial fulfillment of the requirements for the degree of Doctor of Philosophy in Biostatistics for Kevin P. Josey. The content is solely the responsibility of the authors and does not necessarily represent the official views of the National Institutes of Health or the United States Department of Veterans Affairs.

\appendix

\section{Simulation Code}

Code for reproducing the simulation experiment conducted in Section \ref{sim_1} is available at the following address: \texttt{https://github.com/kevjosey/transport-sim}. All data analyzed in this study are publicly available to investigators with approved human subjects approval via the US National Institutes of Health, National Heart, Lung, and Blood Institutes, Biologic Specimen and Data Repository Information Coordinating Center (ACCORD Study, https://biolincc.nhlbi.nih.gov/studies/accord/) or the US National Center for Health Statistics (NHANES, https://www.cdc.gov/nchs/nhanes/). Statistical code for creating analytic datasets and for performing analyses are available from the authors upon request.

\newpage

\bibliography{tbib}

\end{document}